For Table of Contents Use Only

# Nucleation in Sessile Saline Microdroplets: Induction Time Measurement via Deliquescence-Recrystallization Cycling

Ruel Cedeno[1,2], Romain Grossier[1]*, Mehdi Lagaize[1], David Nerini[3], Nadine Candoni[1], Adrian Flood[2]*, Stéphane Veesler[1]*

[1]CNRS, Aix-Marseille University, CINaM (Centre Interdisciplinaire de Nanosciences de Marseille), Campus de Luminy, Case 913, F-13288 Marseille Cedex 09, France

[2]Department of Chemical and Biomolecular Engineering, School of Energy Science and Engineering, Vidyasirimedhi Institute of Science and Technology, Rayong 21210, Thailand

[3] Aix Marseille University, University Toulon, Mediterranean Institute of Oceanography, CNRS INSU,IRD,UM 110, Marseille, France

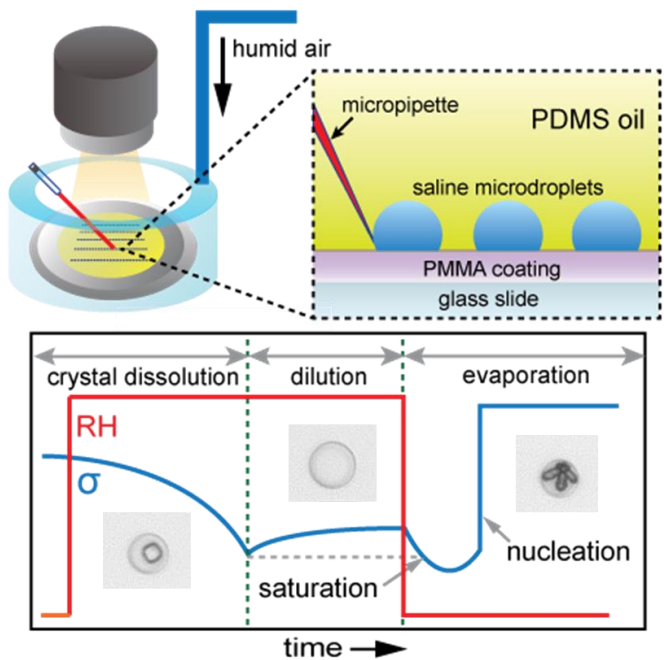

SYNOPSIS: With the aim of quantifying nucleation kinetics, a method for induction time measurement on sessile microdroplets has been developed based on deliquescence-efflorescence cycle and image analysis.

# Nucleation in Sessile Saline Microdroplets: Induction Time Measurement via Deliquescence-Recrystallization Cycling

Ruel Cedeno[1,2], Romain Grossier[1]*, Mehdi Lagaize[1], David Nerini[3], Nadine Candoni[1], Adrian Flood[2]*, Stéphane Veesler[1]*

[1]CNRS, Aix-Marseille University, CINaM (Centre Interdisciplinaire de Nanosciences de Marseille), Campus de Luminy, Case 913, F-13288 Marseille Cedex 09, France

[2]Department of Chemical and Biomolecular Engineering, School of Energy Science and Engineering, Vidyasirimedhi Institute of Science and Technology, Rayong 21210, Thailand

[3] Aix Marseille University, University Toulon, Mediterranean Institute of Oceanography, CNRS INSU,IRD,UM 110, Marseille, France

**ABSTRACT:** Induction time, a measure of how long one will wait for nucleation to occur, is an important parameter in quantifying nucleation kinetics and its underlying mechanisms. Due to the stochastic nature of nucleation, efficient methods for measuring large number of independent induction times are needed to ensure statistical reproducibility. In this work, we present a novel approach for measuring and analyzing induction times in sessile arrays of microdroplets via deliquescence/recrystallization cycling. With the help of a recently developed image analysis protocol, we show that the interfering diffusion-mediated interactions between microdroplets can be eliminated by controlling the relative humidity, thereby ensuring independent nucleation events. Moreover, possible influence of heterogeneities, impurities, and memory effect appear negligible as suggested by our 2-cycle experiment. Further statistical analysis ($k$-sample Anderson-Darling test) reveals that upon identifying possible outliers, the dimensionless induction times obtained from different datasets (microdroplet lines) obey the same distribution and thus can be pooled together to form a much larger dataset. The pooled dataset showed an excellent fit with the Weibull function, giving a mean supersaturation at nucleation of 1.61 and 1.85 for the 60pL and 4pL microdroplet respectively. This confirms the effect of confinement where smaller systems require higher supersaturations to nucleate. Both the experimental method and the data-treatment procedure presented herein offer promising routes in the study of fundamental aspects of nucleation kinetics, particularly confinement effects, and are adaptable to other salts, pharmaceuticals, or biological crystals of interest.

## Introduction

Nucleation in solutions has been a subject of numerous investigations due to its significance in material synthesis[1], pharmaceutical purification[2], and biomineralization[3]. In general, nucleation is the step that determines how long we must wait before the appearance of a stable crystal cluster in a supersaturated solution.[4] This "waiting time", referred to as induction time, is a function of the nucleation rate and the system size. Most induction time measurements are carried out at constant supersaturation for the sake of "simplicity" of data interpretation and modeling[4]. However, in reality, most nucleation processes occur at varying supersaturation, either by cooling, antisolvent, or evaporative crystallization[5]. Thus, a thorough understanding of the nucleation kinetics of such systems is important.

In this context, due to the stochastic nature of nucleation, droplet-based microfluidic systems have been widely used as experimental tool as they permit numerous independent experiments using very small quantities of material. Moreover, their small sizes promote homogeneity in temperature and composition. Despite these advantages, there remain some drawbacks in the use of droplet-based microfluidics in the context of nucleation studies. First, since induction time scales inversely with the system size due to kinetic confinement effect, high supersaturation level must be achieved in order to observe nucleation events in small volumes within reasonable time scales. Consequently, in constant supersaturation experiments, not all of the microdroplets (denoted as μDs) will nucleate, thereby reducing the statistical quality of the data (data is censored). Another issue is that the time for a critical cluster to grow to detectable size depends on the instrument sensitivity, which can also affect the robustness of the measurements[6]. Regarding these limitations, evolving supersaturation experiments where the solvent is allowed to evaporate seem promising[7]: μDs rapidly reach high supersaturations where growth time to detectable size is negligible, and every μD will give rise to a nucleation event (uncensored data). Thus, in our previous work, we have developed an experimental setup allowing facile generation of monodisperse arrays of nanoliter to femtoliter droplets immersed in an oil film, which can serve as evaporative microcrystallizers[8, 9]. We have also shown that a simple and efficient digital-image processing method based on the standard deviation of the grey-level pixels of a single μD and its immediate vicinity (σ) is useful in probing the microdroplet dynamics, particularly the onset of nucleation.[9, 10] Interestingly, we have observed that in these experiments, microdroplets can interact with each other via water diffusion dynamics, i.e. when one µD nucleates, water can diffuse to its closest neighbors. Unfortunately, these interactions would complicate the analysis of induction times obtained from µD arrays.

In this contribution, we develop a new approach to measure induction time in sessile arrays of microdroplets using deliquescence-re-crystallization cycles. With the help of our in-house developed microdroplet generation system with humidity control module, we show that by controlling the relative humidity, the interfering diffusion-mediated interactions between microdroplets can be eliminated. Then, by using appropriate statistical tests, we show that our measurement method is robust and reproducible. Finally, our evaporative nucleation experiments on 60pL and 4pL microdroplets (at equilibrium) confirm the confinement effect[11] where smaller systems reach higher supersaturations.

## Experimental Section

**I. Production of microdroplets with microinjectors** (Fig. 1a). We employed a mechanically controlled micropipette to generate arrays of monodisperse aqueous NaCl sessile microdroplets (1.7 M) on the surface of a poly(methyl methacrylate) (PMMA)-coated glass slide immersed in a 0.4mm-thick layer of polydimethylsiloxane (PDMS) oil (10 cSt). A detailed description of this procedure has been presented previously[8] while the details of the chemical products and equipment used are shown in the Supplementary Information (SI) 1 and 2.

**II. %RH control system**. A home-made, 3D printed (Formlabs printer, PMMA resin) %RH controller module (Fig. 1b), allows us to control the %RH of an air flow on the oil surface, while still having physical access to µDs (not a closed box, but an open-to-air design). Humid air is obtained through consecutive bubbling of a dry air-flow: first in hot-water, then in room-temperature water. A liquid trap ends the process to remove undesired water droplets generated by the bubbling process and conveyed in tubing. Mixing humid-air with dry-air allows fine control of %RH, from a maximum humidity condition (95% RH) to a dry condition (10% RH). The flow is then distributed uniformly above the oil layer. Calibration and stability of %RH were conducted with a %RH sensor (Sensirion SHT85) plugged onto a RaspberryPi3, and a home-made Python3 script to record %RH measurements (see SI3).

**III. ($\sigma$,t) measurements**. The dynamics of each individual µD is followed through an already developed image-processing technique. In a Region-Of-Interest (ROI), centered on each µD and its immediate vicinity as shown in Fig 2, we extract, for each time step (image), a single scalar value: the standard-deviation $\sigma$ of ROI-pixels' grey-levels. This is a post-processing technique: images of the whole µDs-array are acquired at a desired frequency until the end of the experiment, then we process the image-stack with ImageJ (FIJI). First, we extract each µDs edges on first images, and plot a ROI around each one. We then propagate $\sigma$ measurement in each ROI through the whole image-stack to obtain the ($\sigma$,t)-curves which reflects the spatial and temporal dynamics of each µD. ($\sigma$,t)-curves are then processed with home-made Python3 scripts (available on request).

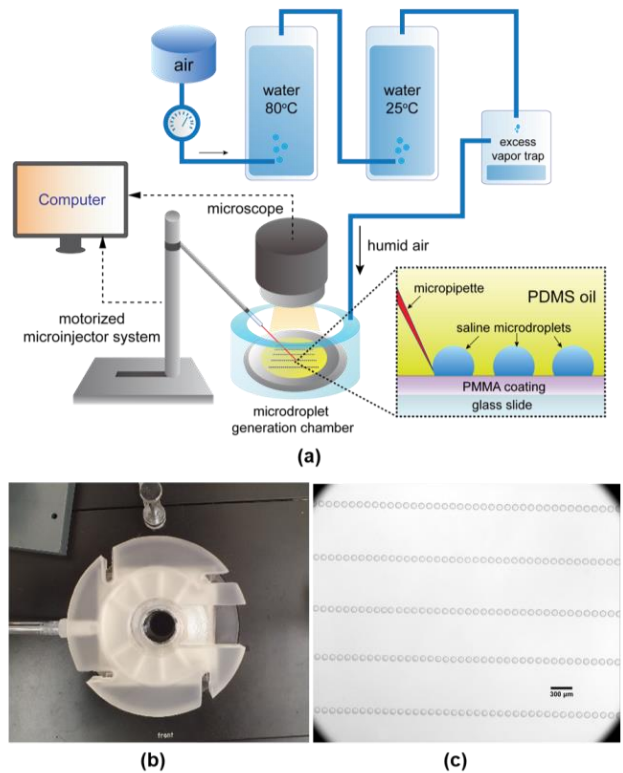

Figure 1. (a) Schematic diagram of the microdroplet generation system with humidity control module (b) picture of the confined microdroplet generation chamber that contains PMMA-coated glass plate immersed in PDMS oil (c) sessile microdroplets printed on the surface of the PMMA using a motorized microinjector.

**IV. Experimentation timeline**. The experiments can be divided into 3 consecutive steps. **Step 0 -** this step is dedicated to obtaining an array of sessile crystals. A µD array is generated (Fig. 1c), following a previously developed method. %RH is here not controlled, and just need to be lower than %RH$_{eq}$, so that µDs will contract (water diffuses to the oil and above-oil atmosphere) until nucleation and growth of a unique crystal in each µD. This step ends when an array of fully-dried crystals is obtained. **Step 1 -** the %RH above the oil layer is increased so that %RH > %RH$_{eq}$ (the %RH corresponding to the equilibrium water partial vapor pressure above a saturated solution). Crystals will then, until complete dissolution, absorb water through the oil-layer. Step 1 ends after every crystal has been completely dissolved and the size of the resulting µDs seems constant. **Step 2 -** the %RH above the oil layer is decreased so that %RH < %RH$_{eq}$. Water then selectively diffuses from µDs, through the oil to the above-oil atmosphere; this contraction ultimately leads to nucleation and growth of a single crystal in each µD. Step 2 ends when we recover the initial conditions of step 1: an array of fully dried sessile crystals. **Cycling -** The sequence Step 1 and Step 2 can then be repeated at will.

## Results and Discussion

The dynamical behavior of the generated µDs depends on the relative humidity %RH above the oil in comparison with the equilibrium %RH$_{eq}$ above the saturated solution (for NaCl in water, %RH$_{eq}$=75% at 25°C). Generally, if %RH > %RH$_{eq}$, the crystal would absorb moisture leading to deliquescence, dissolution, and dilution. If %RH <

%RH$_{eq}$, water would evaporate from the μD leading to an increase in salt concentration until nucleation. By controlling the prevailing %RH, we demonstrate the use of our image analysis protocol to track the dissolution and nucleation of μDs through ($\sigma$,t)-curves.

**I. Analysis of ($\sigma$,t)-curves.** Figure 2 presents the typical time evolution during a nucleation experiment of $\sigma$ measured for each ROI, i.e., for each μD. The analysis of ($\sigma$,t)-curves show characteristic points which are useful data points giving information on the dynamics of each μD, on both dissolution (deliquescence of the crystal), contraction and nucleation (See the video in SI for the whole process). Their sequence and characteristics are discussed hereafter:

**$t_0$ - $t_{DISS}$.** Dissolution time ($t_{DISS}$) is the first encountered characteristic point, appearing as a local minimum on the curve. From $t_0$ (at time t=0) to $t_{DISS}$, the crystal is dissolving through water absorption, and $t_{DISS}$ is the ultimate point of this process, when the crystal was last seen before total dissolution. It is noted that, on the path to $t_{DISS}$, the signal can present large and erratic oscillations, due to small movements of the crystal (jiggling) along the process leading to its total dissolution.

**$t_{DISS}$ - $t_{SHIFT}$.** This is the μD dilution step, where, after the crystal completely dissolves, we allow μDs to continue to absorb water, thereby decreasing their NaCl concentration: μDs are undersaturated. At some time ($t_{SHIFT}$), when μDs do not visually evolve anymore, we shift %RH above the oil-layer, from above %RH$_{eq}$ to a lower value, here from 95% to 10%RH.

**$t_{SHIFT}$ - $t_{MATCH}$.** From the time we shift %RH, each μD will contract by selective diffusion of water through the oil-layer to the controlled atmosphere. Initially, the μDs have a refractive index lower than the oil one, but as contraction occurs, their refractive will increase, and will at some time ($t_{MATCH}$) match the oil refractive index: at this point μDs optically disappear.

**$t_{MATCH}$ - $t_{NUC}$.** The refractive index increases, linked to the increase in NaCl concentration, until crystal nucleation. Therefore, sigma will gradually re-increase after $t_{MATCH}$ and crystal nucleation will appear as a large instantaneous (<1s) jump in sigma values at $t_{NUC}$. The growth of the single nucleated crystal will continue to occur as long as the contraction of the μD continues, until no more water is available: we recover the initial state, a dry sessile crystal.

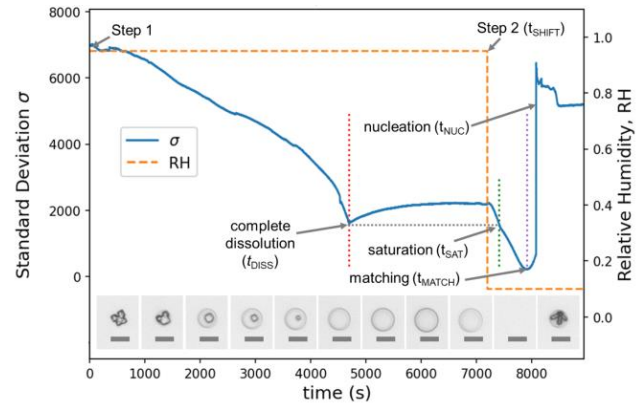

Figure 2. Typical ($\sigma$,t)-curve, during a full %RH cycle (deliquescence/recrystallization), associated to each Region-Of-Interest (ROI). An illustrative ROI is provided in the lower part of the graph to show typical evolution of μDs. Characteristic time points are reported on both %RH curve and ($\sigma$,t)-curve. %RH cycle (dotted orange) goes from step 1 (crystal deliquescence up to $t_{DISS}$ followed by μD dilution) to step 2 (μD contraction) at time $t_{SHIFT}$. During contraction, μD is at saturation at time $t_{SAT}$, optically disappear at time $t_{MATCH}$, and crystals nucleate at time $t_{NUC}$.

**II. Estimating $t_{SAT}$.** For experiments with evolving supersaturation, the time at which the microdroplet becomes saturated ($t_{SAT}$) must be subtracted from the nucleation time $t_{NUC}$ to calculate the induction time.[12] Otherwise, the measured induction time would depend on the arbitrary initial concentration. In principle, $t_{SAT}$ can be estimated by computing the evaporation rate as a function of %RH and system geometry[13] but this could introduce another layer of uncertainty. Alternatively, we can extract $t_{SAT}$ from ($\sigma$,t)-curves by making two assumptions. First, we consider that during the dissolution step of the crystal ($t_0$ to $t_{DISS}$), the solution is in equilibrium with the crystal[14], i.e at saturation concentration. Second, we consider that along the expansion stages ($t_{DISS}$ to $t_{SHIFT}$) and contraction stages ($t_{SHIFT}$ to $t_{MATCH}$), the μD shape have the same symmetric evolution, i.e. its shape follows the same smooth reversible path. Given that $\sigma$ is a function of only shape and refractive index (so NaCl concentration), we can use the value of $\sigma$ at $t_{DISS}$ (where μD is saturated) as reference to find $t_{SAT}$ during the contraction stages (Figure 2). Thus, the μDs are assumed to be at saturation when their $\sigma$ is equal to that of $t_{DISS}$. This is a direct consequence of our symmetric shape path hypothesis.

**III. Eliminating interactions via %RH control.** When the droplet contraction is done at %RH close to the %RH$_{eq}$, the μDs can interact with each other via water diffusion dynamics. This was previously reported and explained in a previous paper.[10] To summarize, when a microdroplet nucleates, it suddenly increases the chemical potential of water inside the microdroplet. As a result, water diffuses from the nucleated microdroplet to its closest neighbors and thereby decreases their solute concentration. This is shown in Figure 3, where selected ($\sigma$,t)-curves show clear oscillations, typical of nucleation induced μD-to-μD interactions. This data was obtained in contracting μDs at 55% RH (room conditions, without using the %RH module). Unfortunately, these interactions would make it impossible to consider each μD as an independent micro-crystallizer. To perform experiments in conditions where one can rule-out any possibility of μD-to-μD interactions, the %RH must be lowered such that the driving force for diffusion towards the atmosphere is high enough to prevent diffusion to neighboring μDs.

For this purpose, we developed two versions of an air-flow dispatcher module (v1 and v2, see SI3). If the interactions are fully eliminated, there should be no oscillations in the σ-curve prior to nucleation in contrast to Figure 3. Although both designs were able to eliminate such oscillations, the v2 module design provides a much better spatial homogeneity of hygrometric conditions as revealed by the statistical-momentums analysis of $t_{DISS}$ distributions (see SI3). This led us to choose the v2 design for the experiments conducted and analyzed here.

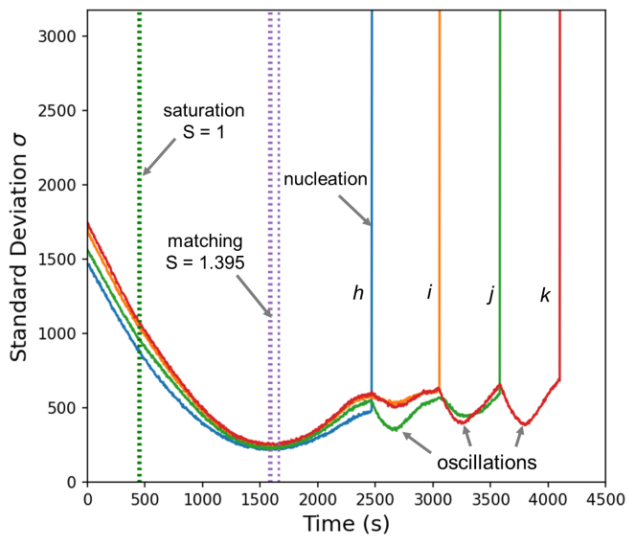

Figure 3. ($\sigma$,t) curves for 4 neighboring µDs, under conditions (room conditions at 55%RH, without using %RH module) which promote µD-to-µD interactions, which appear as oscillations in the $\sigma$ signal. While a µD nucleates a crystal at time 'h', water is released to other neighboring µDs, thus diluting them. Their associated $\sigma$ temporarily decreases before µD contraction re-occurs, thereby re-increasing their associated $\sigma$. Amplitude and time shift of the interactions are driven by both µD and nucleating-µD parameters such as size, supersaturation, and also by their separation distance. The spatial arrangement of µDs, named by their nucleation time, are the following: i, k, j, h. It is notable that these interactions are unobservable on images while looking at µD diameter changes under optical resolution.

**IV. Small and Big-Droplets (SD and BD) experiments.** To study the effect of confinement on nucleation, we performed experiments involving small droplets SD (4 pL) and big droplets BD (60 pL). First, we generated lines of monodisperse aqueous NaCl sessile microdroplets (1.7 M). Big µDs (total number of imaged µDs: 191, in 5 lines: 38, 36, 39, 39 and 39 µDs) were generated first. µDs are left free to contract (**Step 0**), at room conditions (55% RH), until we obtain lines of totally dried sessile crystals. Then, as the %RH above the oil is increased to 95%RH (**Step 1**), we start image acquisitions at a 1Hz rate for BD and 4Hz rate for SD. Following this experiment, we generated smaller µDs (total number of imaged µDs: 170, in 3 lines: 56,56 and 58 µDs), and conducted the same experiment, except that for these small µDs we cycle them twice through Step 1 / Step 2. Images of representative lines of small and large µDs are provided on Figure 4.

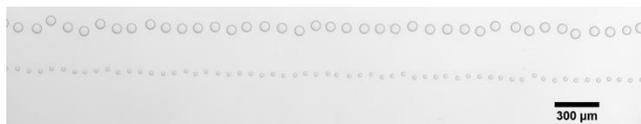

Figure 4. Representative lines of small µDs and large µDs.

**Datasets**. S(t) is the cumulative probability that nucleation has **not** occurred as a function of time t, and is called the survival function. In experiments presented here, the time evolution of the fraction of droplets in which nucleation has not occurred is considered a good estimator of the underlying S(t). Each µDs line is treated as an ensemble (with associated S(t)), and constitute a dataset. We consider datasets relative to the µDs sizes they are associated to: BD datasets (Big µDs) and SD datasets (Small µDs). These datasets are in fact different uncensored (every generated µD give rise to a detected crystal) samplings of the nucleation times survival function, if existing.

**Costs and benefits of increasing supersaturation.** For experimentations conducted at constant supersaturation, there are two major drawbacks, sources of uncertainties in interpreting obtained nucleation data: $t_{GROWTH}$ and data censoring. $t_{GROWTH}$ is the time to grow from the critical size (the nucleation event in itself) to a detectable size, and needs additional hypothesis to be modeled. Data censoring is the fact that not every experiment (supersaturated µDs in our case) will lead to a nucleation event, because nucleation rates may span over a large bandwidth, so that there's missing information. Here, conducted experiments provide access to uncensored data: every single µD leads to a detectable nucleation event, due to constantly increasing supersaturation. For $t_{GROWTH}$, we also benefit from the supersaturation constantly increasing, combined with a kinetic confinement effect[15], as the probability to nucleate scales with the inverse of the volume, µDs reach high supersaturations before nucleation occurs. As a consequence of these high supersaturations, growth from critical cluster to detectable size is fast enough to be ignored: in less than a second, where experiment span over hundreds to thousands, a large crystal is detected. These benefits in experimentations translate in a cost relative to the increased complexity to model nucleation in this evolving supersaturation framework.

**Checking for heterogeneities and memory effect.** A common source of uncertainties in nucleation experiments are heterogeneities, impurities, and their impact on measured nucleation rates (see the discussion on thermal and quenched disorder in reference [4]). The cycling can be viewed as an interesting approach[16] to detect a subset (some µDs in the experiments) where nucleation is mainly due to heterogeneities, which translates in a population of µDs that "statistically always" nucleates first. For SD datasets, that's what Figure 5 is about: with two cycles, we can compare the nucleation ranks of every µD from a cycle to the following. If a population of µDs present heterogeneities leading to faster nucleation rates exist, it could appear as a pattern, for first ranks: they would statistically nucleate faster than other µDs, so in lower rank positions in both cycles. Figure 5 does not show any kind of patterning, the distribution of ranks spread homogeneously in the whole available space. Moreover, it also indicates lack of historical contamination or memory effect from a cycle to the following, that is, nucleation is independent of previous cycle behavior. Nevertheless, further analyses and larger a number of cycles would be needed to confirm this and will be the subject of another study.

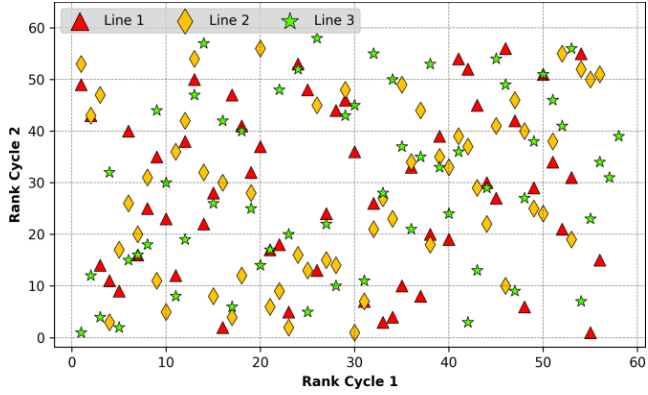

Figure 5. µDs nucleation rank comparison in the two consecutive cycles for SD datasets. "Rank Cycle 1" is µD nucleation rank position in cyle 1 SD datasets, while "Rank Cycle 2" is the µD corresponding rank position it nucleates in cycle 2 SD datasets.

**Distribution of characteristic time points.** Due to the stochastic nature of nucleation, a distribution of times $t_{NUC}$ is expected. But we also see, in Figure 6, that contraction characteristic time points $t_{SAT}$ and $t_{MATCH}$ also distribute: despite optimization in hygrometric conditions, µDs dynamics still show individual characteristics. Such discrepancies can find their source in varying local conditions such as, without being an exhaustive list, surface roughness or chemistry on which µDs rely, local water concentration in oil, small differences or variations in imposed above-oil %RH. Also, small differences in initial diameter, lower than the resolution of our optical system, cannot be ruled out. However, we see in Figure 6 the astonishing similarity of the diverse characteristic time points distributions, in terms of both shape and scale. All µDs lines, whatever the cycle for SDs, shows such resemblance (see SI4). Ultimately, these distributions of $t_{SAT}$ and $t_{MATCH}$ will be reflected in nucleation time $t_{NUC}$ distributions. Also, for the case of SDs, which experienced two deliquesence/nucleation cycles, Figure 6 shows a **Cycle-to-Cycle times location shift** in distributions of times $t_{MATCH}$ and $t_{NUC}$, similar for every SD lines. It seems that, statistically, µDs are slower to reach $t_{MATCH}$ and $t_{NUC}$ during cycle 2 as compared to cycle 1: µDs contract slower in cycle 2.

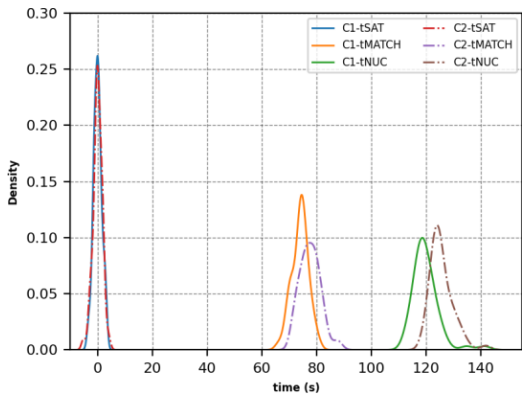

Figure 6. Distributions (kernel density estimation of probability density function) of characteristic time points ($t_{SAT}$, $t_{MATCH}$ and $t_{NUC}$), for a typical dataset (here SD line 1 dataset), for both deliquesence/nucleation cycles (C1 for cycle 1, C2 for cycle 2). Plots are against a shifted time where the mean $t_{SAT}$, respectively to the cycle, is taken as the origin of times.

**Dimensionless Induction Time τ (DIT).** To compare µDs and associated $t_{NUC}$, we have to take into account differences that may only arise from their individual behavior (discrepancies in contracting dynamics), and also from their "cycle" behavior (the location shift shown from cycle 1 to cycle 2 for both $t_{MATCH}$ and $t_{NUC}$ distributions). For this purpose, we non-dimensionalized individual induction times $t_{NUC}$ relative to $t_{SAT}$ and $t_{MATCH}$. The individual µDs times to nucleation $t_{NUC}$ are converted to a dimensionless induction time, as specified by equation 1, here:

$$\tau = \frac{t_{NUC} - t_{SAT}}{t_{MATCH} - t_{SAT}} \qquad (1)$$

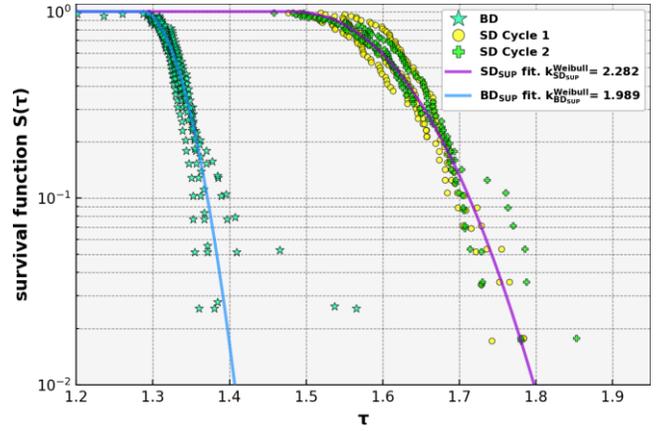

Figure 7. Survival function (sf) semi logarithmic plot of Dimensionless Induction Times (DITs) of every µD line in the full dataset of big droplets (BD) and small droplets (SD). Among SD datasets we discriminate Cycle 1 (SD Cycle1) from Cycle 2 (SD Cycle2). Best fits from a Weibull sf are here provided as guide to the eye, calculated for the full pooling of all BD datasets (BD fit.), and for a subset (all pooled but line 3 in cycle 1) of the full pooling of SD datasets (SD fit.). Associated respective Weibull shape parameters are reported as $k_{BD}^{Weibull}$ and $k_{SD}^{Weibull}$.

**DITs seem similar.** DITs semi logarithmic plots of their survival function are reported on Figure 7. At first sight, such nucleation dimensionless time allows comparison of µDs with individual contraction dynamics, which first benefit, for the SDs case, is to make cycle-to-cycle $t_{NUC}$ location shifts vanish: there is no evidence of a location shift (see SI5), from cycle 1 to cycle 2, so that nucleation times as reported through the use of DITs compare well. It seems, to the eye, that lines in BD datasets and SD datasets can, respectively, be fully pooled to be treated as representative of a unique experiment, thereby increasing the statistically representation. This will be addressed hereafter in the "**Datasets pooling**" section. Also evident on Figure 7, for both BD and SD datasets, the log(sf) of their respective DITs do not follow a purely exponential decay, which would appear as a straight line on the plot: lowest values of DITs draw a plateau. This plateau is both typical of experimentation inaccuracies ($t_{GROWTH}$, heterogeneities), and of compressed exponential decays. As a reason for this plateau, we can here eliminate inaccuracies that arise from a shift between nucleation and detection of the crystal as growth is almost instantaneous to a visible and consequent size: growth time does not pollute these measurements. Also, as shown on Figure 5, the lack of patterns (at least for SDs, the only µDs that underwent two cycles) in their cycle-to-cycle ranks do not promote evidence of a µD population (which nucleates first) subject to a higher nucleation rate due to nucleation promoting heterogeneities. In doing so, this plateau seems to have an existence independently of experimentations inaccuracies, and will be treated as such, promoting **datasets fittings** (addressed hereafter in a specific part) with the help of a typical stretch/compressed distribution: the Weibull distribution, which shape parameter k discriminate between stretching (k<1) and compression (k>1) of an exponential decay. Weibull survival function, in its standardized form, can be written as follows (eq.2), with k being the shape parameter:

$$S(t) = \exp(-t^k) \qquad (2)$$

**Datasets pooling.** Up to now, we implicitly considered a µD line as the scale of the sampling of the nucleation times distribution, each line forming a dataset, and kept comparing line-to-line similarities. There is high temptation to pool all datasets toward a "superdataset" that could be treated as a whole, but the problem would be to justify such data pooling. Such authorization for datasets pooling could rely on the use of a non-parametric statistical test: the "**k-sample Anderson-Darling test**" (**kADt**).[17] The kADt tests the null hypothesis "H0: the k-samples are drawn from the same population" versus the alternative hypothesis "H1: the k-samples are drawn from different populations". In other words, do here presented DITs associated to each µD lines could be considered as samples representative of the same distribution, i.e., could we model these samples with the same distribution function, and pool datasets? For the 5 BD datasets (see SI6), with a 5% significance level, kADt fails to reject the null hypothesis H0: the 5 µD lines and associated DITs distributions can be modeled with the same distribution function, and we pool datasets in $BD_{SUP}$ superdataset. Regarding (see SI6) the 6 SD datasets (considering each cycle - 3 lines each - as independent sampling as lines could be), the kADt reject null hypothesis H0: the 6 datasets cannot be considered to be drawn from the same distribution. We then test every combination of 5 among 6 lines with kADt, and one SD dataset seems to be the source of the null hypothesis rejection: line 3 in cycle 1. In removing this dataset, kADt cannot reject the null hypothesis H0 anymore, and we can then construct a SD superdataset $SD_{SUP}$ constituted of 5 pooled SD datasets: lines 1 and 2 in cycle 1, and every 3 lines in cycle 2.

**Datasets Fittings.** Complementary to the analysis of Survival function (sf) semi logarithmic plots of Dimensionless Induction Times (DITs) presented on Figure 7, the contracting process (where, as volume decreases, supersaturation increases, imposing also nucleation rate to do so upon nucleation of a crystal) encompassed by µDs also justify the modeling of DITs with a Weibull distribution function, which would correspond to a compressed exponential function of class III as suggested by Sear.[4] A fitting procedure was applied to newly constructed $SD_{SUP}$ and $BD_{SUP}$ datasets, and results are presented on Figure 8. It is to note that sensitivity of fittings algorithm to both outliers and initial fitting guesses was addressed in two steps: we first fit "roughly", with guessed outliers removed, then we re-incorporate 98% of the full dataset to refine fit. $R^2$ and mean squared errors were calculated on the full dataset. Despite such precautions, as seen on Figure 8, there is departure of the fit from data at distribution tail (highest DITs). This is not an artefact, tail is not well fitted, but using semi logarithmic scale here does not render justice to quality of fittings, as measured with both $R^2$ and mean squared errors (see SI7). Moreover, Figure 8 confirms confinement effects where the induction time shifts due to system (µD) size: the smaller the system, the longer the induction time.

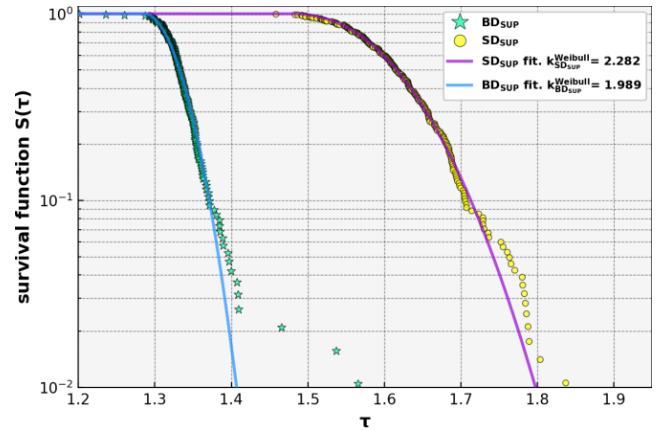

Figure 8 Survival function (sf) semi logarithmic plot of Dimensionless Induction Times (DITs) for both $BD_{SUP}$ and $SD_{SUP}$ datasets. Best fits with a Weibull function are shown as '$SD_{SUP}$ fit.' and '$BD_{SUP}$ fit.' With reported associated distribution shape parameters as $k_{BD}^{Weibull}$ and $k_{SD}^{Weibull}$.

**Supersaturations at nucleation times.** To translate DITs in supersaturations, one need a contraction model: how µDs volume evolves with time. An exact and precise model is complex and out of the scope of this paper, and will be the object of specific studies later on. But from previous works, and for the sake of simplicity, we can choose a linear evolution of volume with time, which allow (see SI8) to calculate supersaturations $\beta_{NUC}$ from DITs and supersaturation $\beta_{MATCH}$ at matching time $t_{MATCH}$, following equation here:

$$\beta_{NUC} = \frac{1}{1 - \tau \times \left(1 - \frac{1}{\beta_{MATCH}}\right)} \qquad (3)$$

In Figure 9, we report calculated supersaturation distributions for both $BD_{sup}$ and $SD_{sup}$ superdatasets, with corresponding Weibull function best fits for both. For $BD_{SUP}$, nucleation occurs at a mean supersaturation of 1.61 and 95% spans from 1.58 to 1.65. For $SD_{SUP}$, nucleation occurs at a mean supersaturation of 1.85 and 95% spans from 1.74 to 2. There is high correspondence with supersaturation values already reported in literature, for comparable system sizes[18-20]

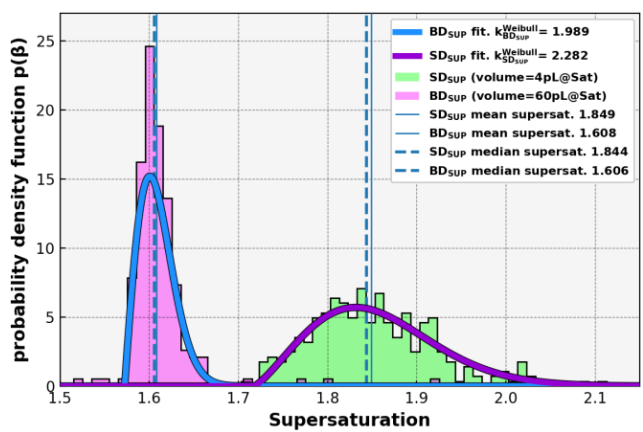

Figure 9. Supersaturation histograms and best fits with a Weibull probability density function (**pdf**), in using a linear decrease of droplets volume with time model. SD and BD are the pooled data as determined by Anderson-Darling k-sample tests. 'SD fit.' and 'BD fit.' are, respectively, SD and BD Weibull **pdf** best fits. Median and mean times of each fits are reported on graph for both SDs and BDs fits.

## Conclusions

In this communication, we present a microdroplet generation system with humidity control module and a novel approach for measuring and analyzing induction times in lines of saline sessile microdroplets undergoing a deliquescence/recrystallization cycling process. With the help of a recently developed image analysis protocol, we first assert microdroplets act independently: there is no microdroplet-to-microdroplet interactions. Second, using a dimensionless time, dynamics of microdroplets along deliquescence/recrystallization cycles are specified, allowing to obtain precise estimation of the nucleation induction times survival function S(t), for every microdroplet lines. Considered as samples of the underlying (unknown) nucleation probability density function, line datasets that satisfies a k-sample Anderson-Darling test are pooled, and resultant superdatasets are fitted with a Weibull function. Moreover, evaporation experiments on 60pL and 4pL microdroplets, at saturation confirm confinement effect that smaller systems reach higher supersaturations. Both the experimental method and the data-treatment procedure seems promising for the study of fundamental aspects of nucleation kinetics, confinement effects in particular, and is adaptable to other salts, pharmaceuticals, or biological crystals of interest.

The modeling of evaporation and the use of a modified Poisson distribution which considers the time-dependence of the nucleation driving force[5] will be a subject of a future work.

## Supporting Information

[Video of the whole process](#) during a full %RH cycle (deliquescence/recrystallization)
Details of Chemical Products
Details of Instrumentation
Humidification control and dispatching module
Distribution of sigma-curves characteristic points versus time
Cycle-to-cycle nucleation times shift.
Anderson-Darling k-sample test for different aggregation subsets.
Modeling of DITs with a Weibull distribution function
Supersaturation Calculation

## Author Information

### Corresponding Authors

A.F., R.G. and S.V.

### Author Contributions

Conducted the experiments and collected data: R. C.

Designed experiments: all but D. N.

Analyzed datasets: all but M. L.

### Acknowledgement

R. Cedeno acknowledges the financial support of Vidyasirimedhi Institute of Science and Technology (VISTEC) and the Eiffel Excellence Scholarship (N°P744524E) granted by the French Government.

### Abbrevations

SD: Small Droplets

BD: Big Droplets

µD: microdroplet

# Nucleation in Sessile Saline Microdroplets: Induction Time Measurement via Deliquescence-Recrystallization Cycling

Ruel Cedeno[1,2], Romain Grossier[1]*, Mehdi Lagaize[1], David Nerini[3], Nadine Candoni[1], Adrian Flood[2]*, Stéphane Veesler[1]*

[1]CNRS, Aix-Marseille University, CINaM (Centre Interdisciplinaire de Nanosciences de Marseille), Campus de Luminy, Case 913, F-13288 Marseille Cedex 09, France

[2]Department of Chemical and Biomolecular Engineering, School of Energy Science and Engineering, Vidyasirimedhi Institute of Science and Technology, Rayong 21210, Thailand

[3] Aix Marseille University, University Toulon, Mediterranean Institute of Oceanography, CNRS INSU,IRD,UM 110, Marseille, France

**SI1**
**Details of Chemical Products**

| Product | Vendor | Properties |
|---|---|---|
| Sodium chloride, NaCl | R.P Normapur ® | Purity = 99.5%<br>Refractive index = 1.5442 |
| Polymethylmethacrylate, PMMA | ALLRESIST GmbH | Molecular weight= 950,000 g/mol<br>Refractive index = 1.395 |
| Polydimethylsiloxane, PDMS oil | Alfa Aesar | Molecular weight = 1250 g/mol<br>Viscosity = 10 cSt<br>Refractive index = 1.3990 |
| Ultrapure water | via Milli-Q Purifier | resistivity = 18.2 MΩ·cm<br>TOC value < 5 ppb |

**SI2**
**Details of Instrumentation**
To avoid microdroplet spreading and coalescence, we coated the glass cover slip with a hydrophobic PMMA resin. For this, glass coverslips (18-mm diameter, cleaned via plasma treatment) were spincoated at 4000 rpm for 1 min (SPIN 150, SPS) with PMMA which were then annealed for 10 min at 170°C. The coverslips were then covered with a 0.8 mm thick layer of PDMS oil. The saline microdroplets were generated on the cover slip by a micropipette with an internal diameter of 0.5 um (Femtotip Eppendorf). The micropipette is mechanically controlled by a home-made motorized micromanipulator consisting of 3 miniature translation stages (piezo electric, MS30 Mechonics) which allows displacement of the micropipette holder in three dimensions by steps of 16 nm. A series of 16-bit images were obtained using an optical microscope (Zeiss Axio Observer D1 equipped with an ANDOR neo sCMOS camera). Images were processed using FIJI software (Image J, NIH, USA) which calculates σ for each region containing microdroplets.

## SI3
## Humidification control and dispatching module

The humidification chamber (airflow module dispatcher) was custom fabricated via SLA 3D-printing (Formlabs, clear resin).

## Improving Spatial Homogeneity

To choose airflow module dispatchers, we compared measured distributions of tDISS in both models (v1 and v2).

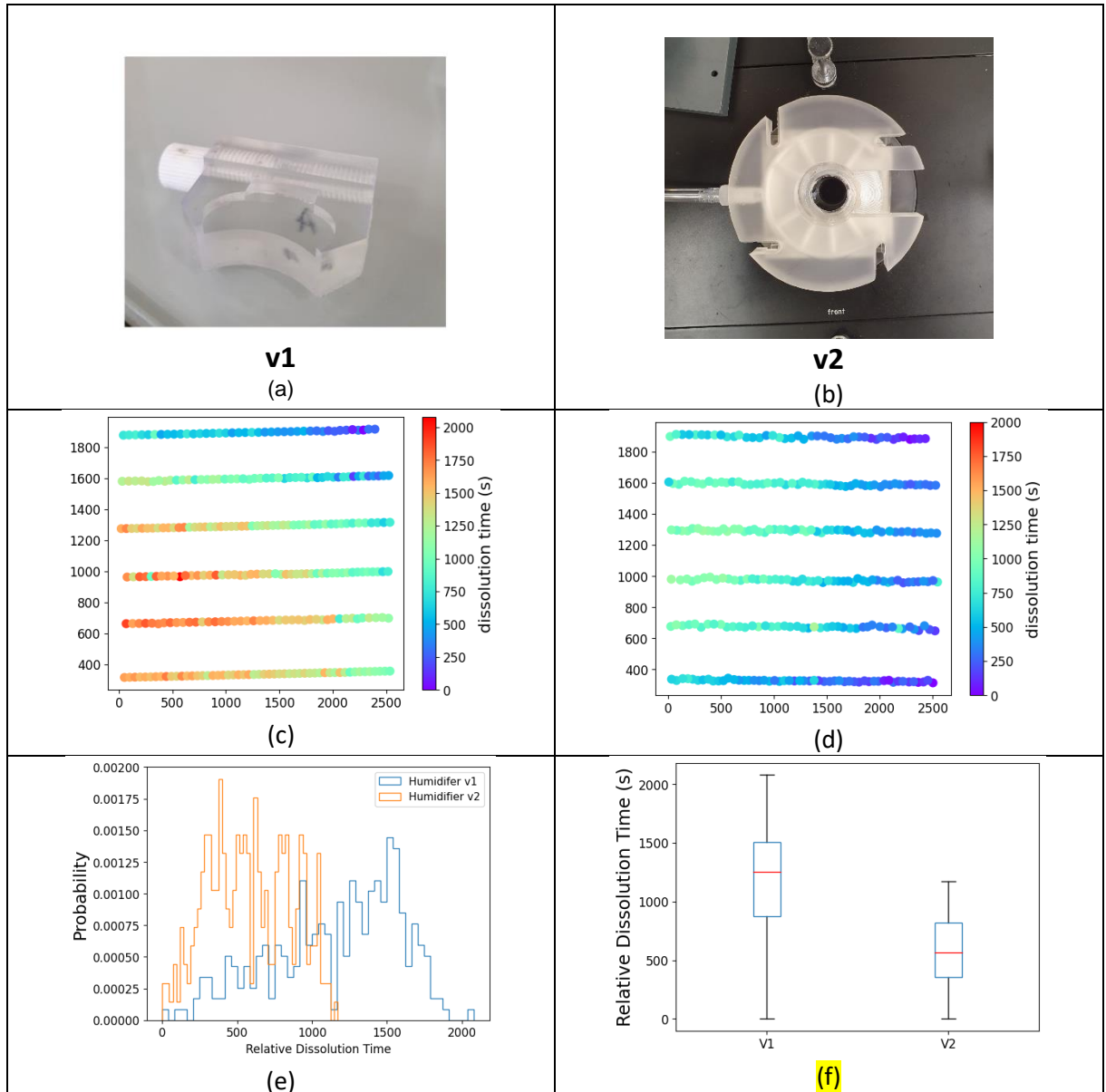

|  | V1 | V2 |
|---|---|---|
| mean | 1162 | 584.0 |
| standard deviation | 426 | 275 |
| skewness | -0.516 | 0.0448 |
| kurtosis | -0.599 | -0.976 |

We have significantly improved the spatial homogeneity by making the humidifier system more symmetric. This is illustrated in the map of dissolution times (c,d), histogram and the box plot (e,f), and the values of mean, standard deviation, skewness and kurtosis (Note that the minimum dissolution time is set to zero).

With this v2 design, to assess asses the speed at which the humidity can be changed, we measured the %RH in the microdroplet generation chamber with "dry" air (directly obtained from compressed air pipelines) and our humid air.

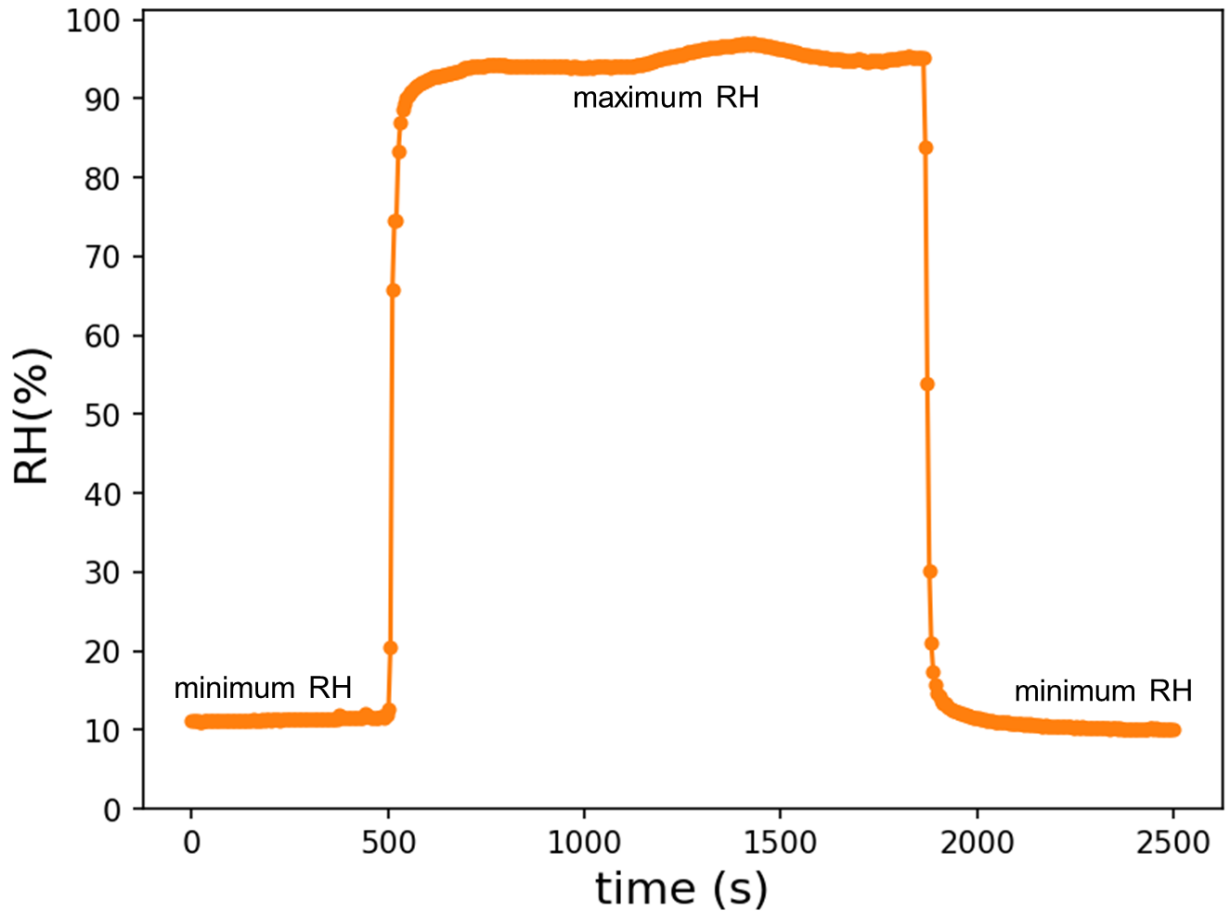

**Figure S1** Relative humidity vs time in the humidifier when RH is shifted from minimum to maximum value and vice versa. This suggests that our humidity control system can almost instantaneously change the RH of our microdroplet generation chamber (negligible lag period). We also show we can maintain a reasonably stable RH as needed.

## SI4
## Distribution of sigma-curves characteristic points versus time

Distribution of sigma-curves characteristic points versus time for BDs lines.

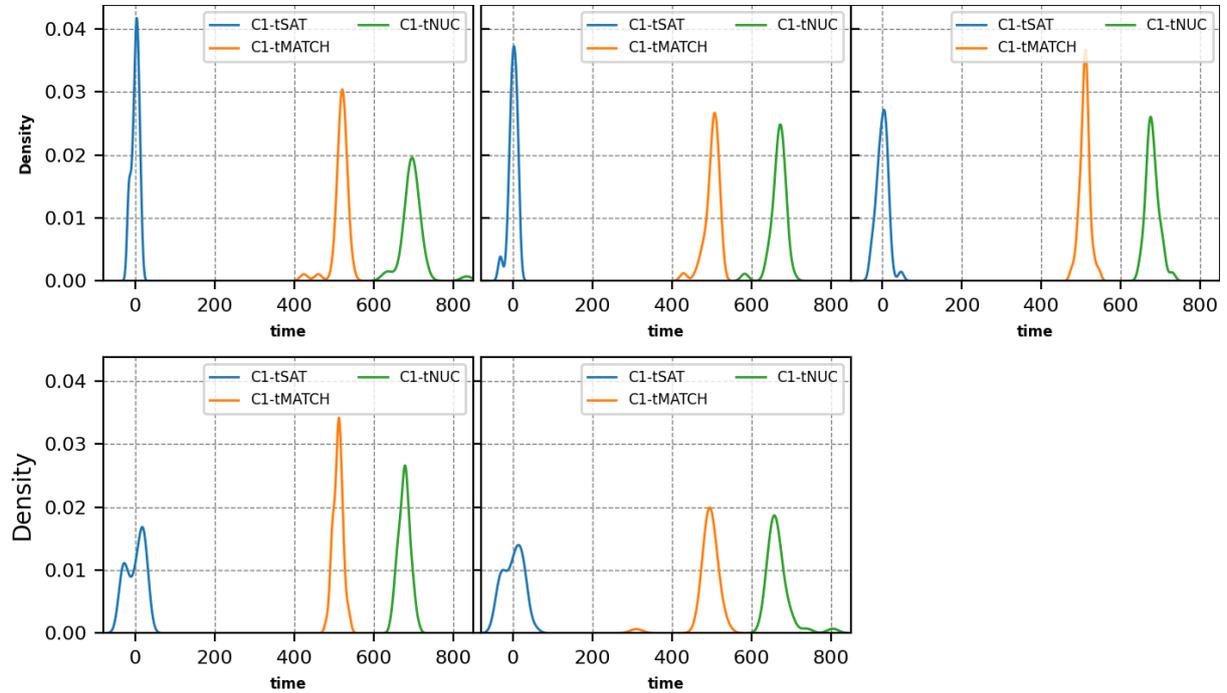

Distribution of sigma-curves characteristic points versus time for SDs lines.

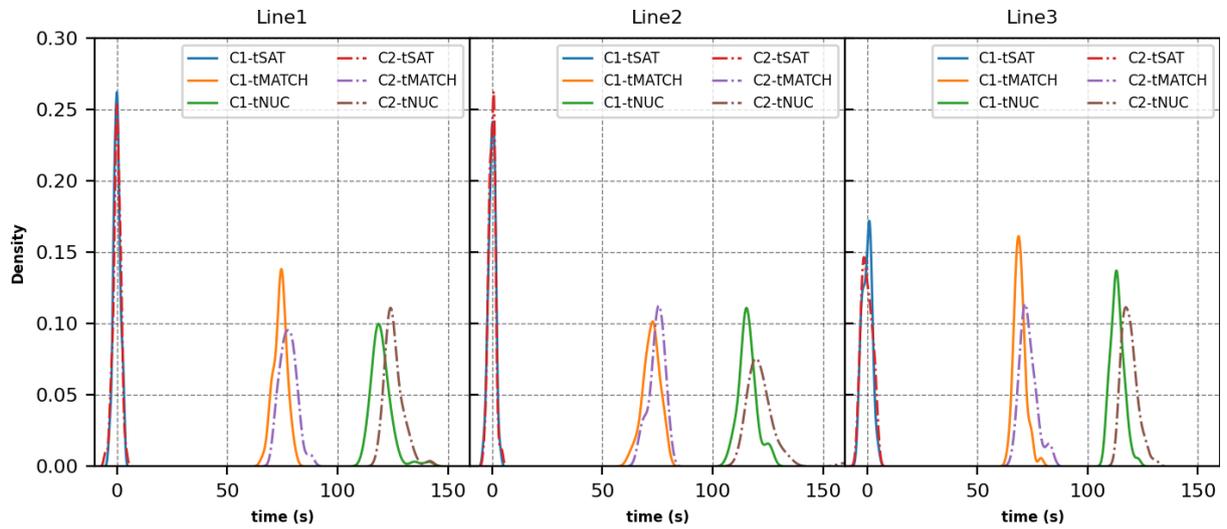

**SI5**
**Cycle-to-cycle nucleation times shift.**
Distribution, for SDs datasets, of dimensionless induction times, while cycle-aggregated. There is no evidence of cycle shifting.

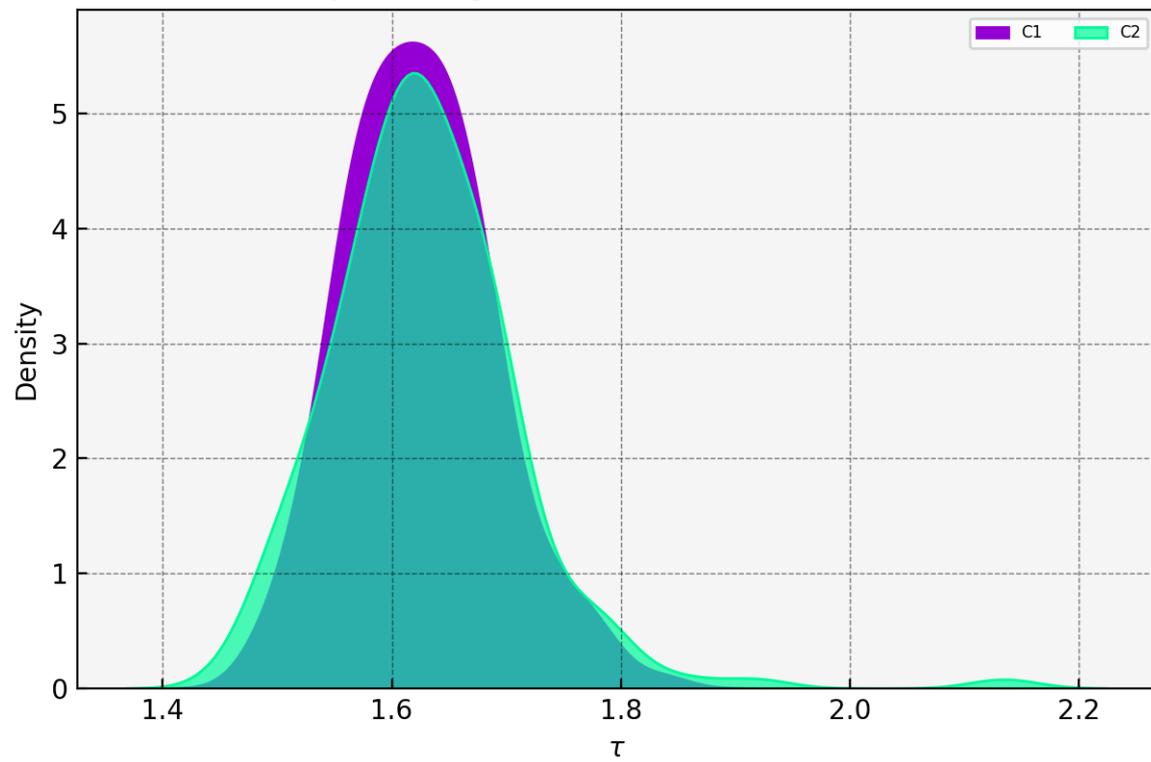

## SI6
## Anderson-Darling k-sample test for different aggregation subsets.

Anderson-Darling k-sample test tests for:
- "null hypothesis **H0**: samples data are drown from the same distribution"
- "alternative hypothesis **H1**: samples data are drawn from different distributions".

We fix risk to reject H0 despite true at a usual value of p-value=.05.

BD: Big Droplets datasets (line number)
k=5
(1, 2, 3, 4, 5) -> 1.7277 / 0.0606         Ok
The pooling of every BD datasets succeeds.

SD: Small Droplets datasets (SD_cycle_line format, ex: SD23 is line 3 in cycle 2)
K=6
['SD11', 'SD12', 'SD13', 'SD21', 'SD22', 'SD23'] -> 2.5189 / 0.0212     X
The pooling of every datasets fail.
We can then try different combinations of 5 among 6 datasets.
K=5
['SD11', 'SD12', 'SD13', 'SD21', 'SD22'] -> 2.6507 / 0.0191         X
['SD11', 'SD12', 'SD13', 'SD21', 'SD23'] -> 2.9413 / 0.0131         X
['SD11', 'SD12', 'SD13', 'SD22', 'SD23'] -> 3.2549 / 0.0087         X
['SD11', 'SD12', 'SD21', 'SD22', 'SD23'] -> 0.6200 / 0.2275         Ok
['SD11', 'SD13', 'SD21', 'SD22', 'SD23'] -> 1.3282 / 0.0984         Ok
['SD12', 'SD13', 'SD21', 'SD22', 'SD23'] -> 2.7799 / 0.0162         X
There are combinations which succeed test, we choose the highest statistical significance (0.2275).

# SI7
# Modeling of DITs with a Weibull distribution function

$$S(x) = exp(-x^k)$$

With $x = \frac{\tau - location}{scale}$

| DATA | k | Location | Scale | R2 | MSE |
|---|---|---|---|---|---|
| $SD_{SUP}$ | 2.281658 | 1.475360 | 0.165096 | 0.998777 | 0.000105 |
| $BD_{SUP}$ | 1.989330 | 1.285815 | 0.056227 | 0.994003 | 0.000477 |

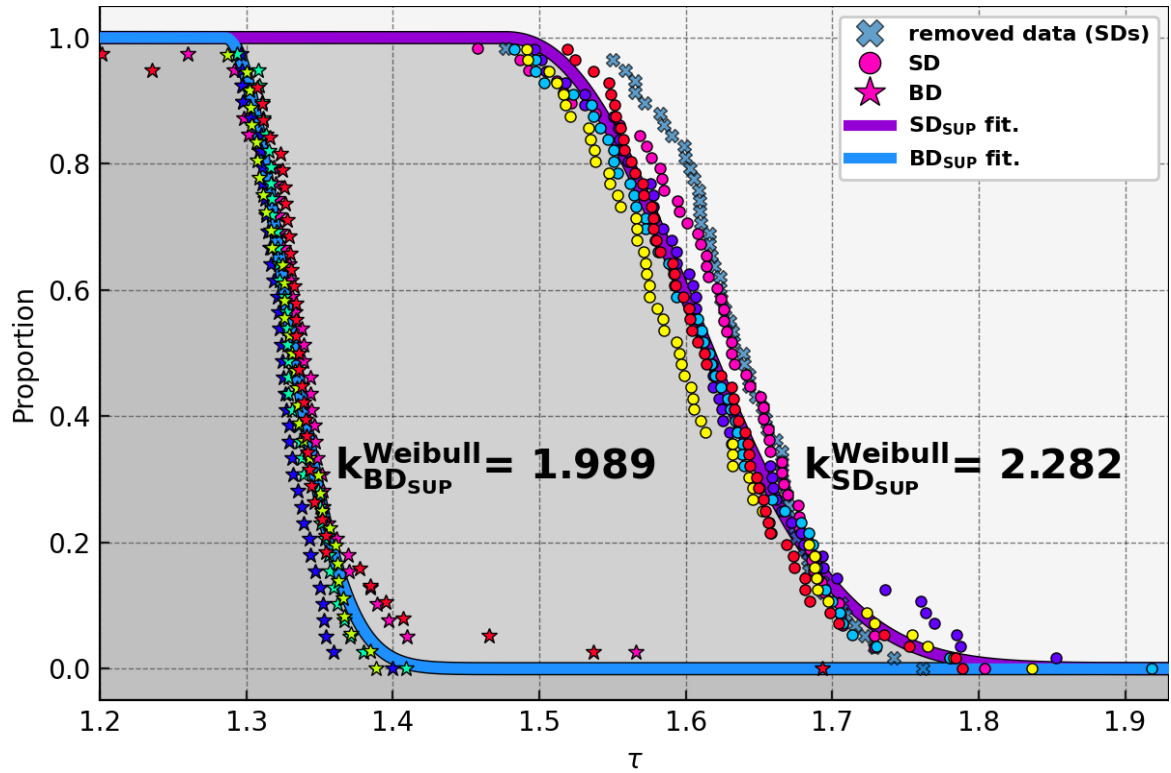

Figure 8. Survival function (sf) plot of Dimensionless Induction Times τ (DITs). Supersaturation **sf** data for SD and BD droplets are superimposed on respective best fits 'SD fit.' And 'BD fit.'. 'removed data (SDs)' is the data of the only line in SD datasets removed by Anderson-Darling k-sample test: line 3 in cycle 1.

**SI8
Supersaturation Calculation from DITs
Determination of Supersaturation Ratio at Refractive Index Matching**

The refractive index of NaCl solution at ambient conditions is plotted against supersaturation ratio. The supersaturation ratio that corresponds to the refractive index of PDMS oil is the supersaturation ratio during refractive index matching ($S_m$). Data were taken from https://www.topac.com/Salinity_brix.html.

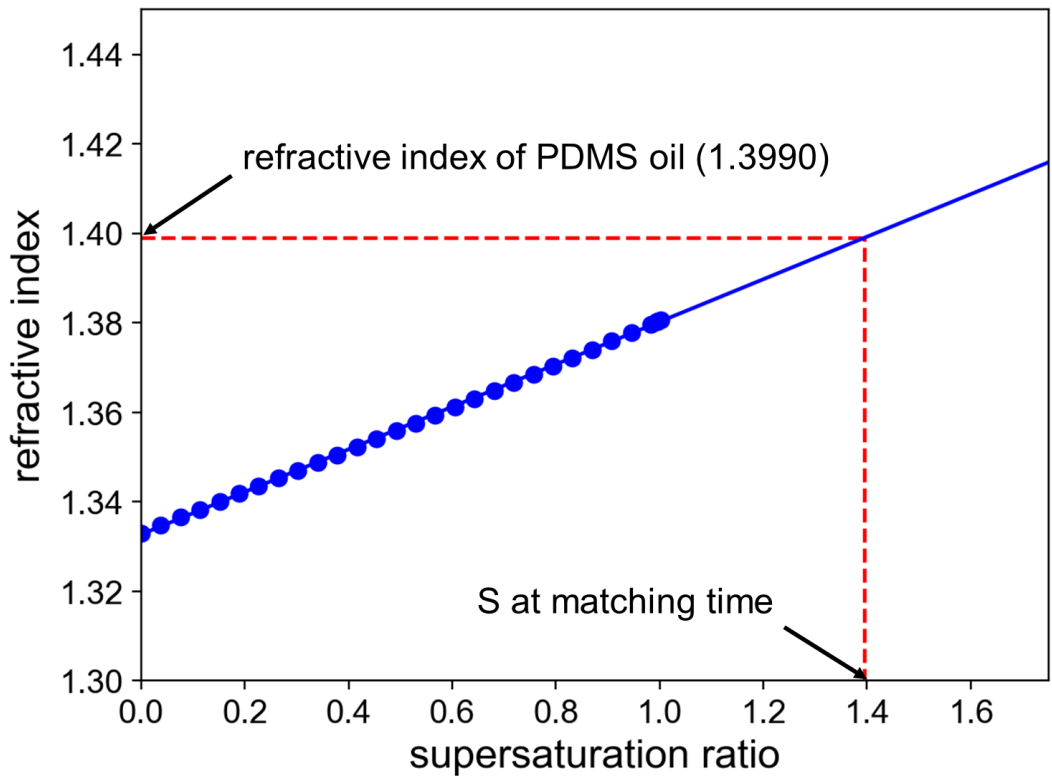

**Figure S2** Refractive index as a function of supersaturation ratio. The refractive index of the droplet matches that of the PDMS oil at S = 1.3990.

From the hypothesized linear volume decrease with time (coefficient alpha), we have, for Vsat the droplet volume at saturation:

V(t) = Vsat (1- alpha*(t - tsat))      eq1

With mass conservation on dissolved salt: S(t)V(t)= cste
Equation 1 can be expressed in terms of supersaturation (with Ssat=1)
1/S(t) = 1 - alpha*(t - tsat)     eq2

If we express volume decrease between tsat and tmatch, we can extract alpha:

Alpha = (1 - 1/Smatch) / (tmatch - tsat)

Knowing alpha, and expressing equation 2 at dimensionless nucleation time tau

Snucleation = (1 - tau*(1-1/Smatch))^-1